\documentclass[ journal ]{IEEEtran}
\usepackage[T1]{fontenc}

\ifCLASSINFOpdf
\else
\usepackage[dvips]{graphicx}
\fi
\usepackage{url}

\usepackage{amsmath}
\usepackage{amsthm}
\usepackage[cmintegrals]{newtxmath}

\usepackage[T1]{fontenc}
\usepackage{amsmath}
\usepackage{amsthm}
\usepackage[cmintegrals]{newtxmath}
\usepackage{amssymb}
\usepackage{bm}
\usepackage{bbm}
\usepackage{comment}
\usepackage{graphicx}
\usepackage{calc}
\usepackage{epstopdf}

\newtheorem{proposition}{Proposition}

\usepackage{verbatim}
\usepackage{color}
\usepackage{array}
\usepackage{cite}
\usepackage{stfloats}

\usepackage{xcolor}
\usepackage{floatrow}
\usepackage{lipsum}
\usepackage[ruled,norelsize]{algorithm2e}
\usepackage{algorithmic}
\usepackage[caption=false,subrefformat=parens,labelformat=parens]{subfig}
\usepackage{multirow}
  \usepackage[a4paper, left=0.6 in, right=0.6 in, top=0.7in, bottom=1.1 in ]{geometry}

\begin{document}
\title{ Decentralized Beamforming Design  for  Intelligent Reflecting Surface-enhanced Cell-free Networks}

		
\author{
	Shaocheng~Huang, Yu~Ye, \IEEEmembership{Student Member, IEEE}, Ming~Xiao, \IEEEmembership{Senior Member, IEEE}, H. Vincent Poor, \IEEEmembership{Fellow, IEEE}, and Mikael Skoglund, \IEEEmembership{Fellow, IEEE}
	\thanks{This work was supported in part by the U.S. National Science Foundation under Grant CCF-1908308.}
	\thanks{S.~Huang, Y.~Ye, M.~Xiao and M.~Skoglund are with the  school of Electrical Engineering and Computer Science, KTH Royal Institute of Technology, Stockholm, Sweden (email: \{shahua, yu9, mingx, skoglund\}@kth.se).}
	\thanks{ H. V. Poor is with the Department of Electrical Engineering, Princeton University, Princeton,  USA (email: poor@princeton.edu).}
}  

\maketitle
 
\begin{abstract}
Cell-free networks are considered as  a promising distributed network architecture to  satisfy the increasing number of users  and high rate expectations in beyond-5G systems. However, to further enhance   network capacity, an increasing number of high-cost base stations (BSs) is required. To address this problem and inspired by the cost-effective intelligent reflecting surface (IRS) technique, we propose a fully decentralized design framework for cooperative beamforming  in IRS-aided cell-free networks.  We first transform the centralized weighted sum-rate maximization problem into a tractable consensus optimization problem, and then an incremental alternating direction method of multipliers (ADMM) algorithm is proposed to locally update the beamformer. The complexity and convergence of the proposed method are analyzed, and these results show that the performance of the new scheme can asymptotically approach   that of the  centralized one as the number of iterations increases. Results also show that IRSs can significantly increase the system sum-rate of cell-free networks and the proposed method outperforms existing decentralized methods.
\end{abstract}

\IEEEpeerreviewmaketitle

\section{Introduction}

Recently, a user-centric network paradigm called cell-free networks has been considered as a promising technique to provide high network capacity and overcome the cell-boundary effect of traditional network-centric networks (e.g., cellular networks)\cite{buzzi2017cell,mai2020downlink,ngo2017cell,interdonato2020local}.  In cell-free networks, a large number of distributed service antennas, 
which are connected to central processing units (CPUs), coherently serve all users on the same time-frequency resource\cite{mai2020downlink}. This distributed communication network can offer many degrees of freedom and high multiplexing gain.  
 Recent results show that cell-free networks outperform traditional cellular and small-cell networks in several practical scenarios\cite{ngo2017cell,mai2020downlink}.  
To provide high directional gains, beamforming design is important in cell-free networks. To cooperatively design beamforming, a centralized zero-forcing (ZF) beamforming scheme is proposed in \cite{nayebi2017precoding}. Since the CPU should collect all instantaneous channel state information (CSI) of all base stations (BSs), centralized approaches might be unsalable when the number of BSs and users (UEs) is large and the beamforming optimization at the CPU may be overwhelming due to the high dimensionality of aggregated beamformers. To avoid instantaneous  CSI exchange among BSs via backhauling and reduce  computation complexity at CPU, most recent works assume a simple non-cooperative beamforming strategy at the BSs, e.g., maximum ratio transmission (MRT)\cite{ngo2017cell} and local ZF \cite{interdonato2020local}. However, cooperation among BSs is not considered, and thus  interference among BSs cannot be efficiently eliminated. 
 Though a distributed beamforming scheme is introduced in \cite{atzeni2020distributed}, it is not fully decentralized and each local update requires extensive CSI exchange among   BSs.


To further increase the capacity of cell-free networks, the deployment of more distributed BSs requires high hardware cost and power consumption\cite{ngo2017cell}. Moreover, when a cell-free network implemented at high-frequency bands (e.g., millimeter-wave bands), it might suffer severe propagation loss and be vulnerable to blockage\cite{rappaport2017overview}. Meanwhile, an emerging technique called intelligent reflecting surface (IRS) equipped with low-cost, energy-efficient and high-gain meta-surfaces can potentially address the above  problems\cite{bjornson2019intelligent,guo2020weighted,zhang2020joint}. In \cite{bjornson2019intelligent}, it is shown that the IRS outperforms decode-and-forward relaying if the size of the IRS is large. In addition, a centralized beamforming scheme of cell-free networks is proposed in \cite{zhang2020joint}, in which a part of   BSs in the network is replaced by IRSs to improve the network capacity at low cost and power consumption. It is shown that the cell-free network with IRSs can achieve a larger weighted sum-rate (WSR) than that without IRSs. However, there is no decentralized beamforming scheme for  IRS-aided cell-free networks.    

Based on above observations,  we propose a fully decentralized design framework for cooperative beamforming in IRS-aided cell-free networks, in which transmitting digital beamformers and IRS-based analog beamformers are jointly optimized. Specifically, according to fractional programming (FP), we first transform the centralized beamforming optimization problem into a tractable  consensus problem for decentralized optimization. Then, based on the alternating direction method of multipliers (ADMM), a fully decentralized beamforming scheme is proposed to incrementally and locally update the beamformers. Since only three variables are incrementally updated and transmitted to the next BS, our scheme can significantly reduce backhaul signaling compared with full CSI exchange among  BSs. Moreover, we use a low-complexity majorization-minimization (MM) method to efficiently optimize the IRS-based analog beamformer with non-convex constraints. Additionally, since the reflection element only has  finite reflection levels in practice, we then optimize the IRS-based analog beamformer with low-resolution phase shifts.  Finally, the convergence of the proposed decentralized scheme is proved and computation complexity is analyzed.

%
%


 \section{System model}  
\begin{figure}[b] 
    	\vskip 0in
    	\begin{center}
    		\centerline{\includegraphics[width=78mm]{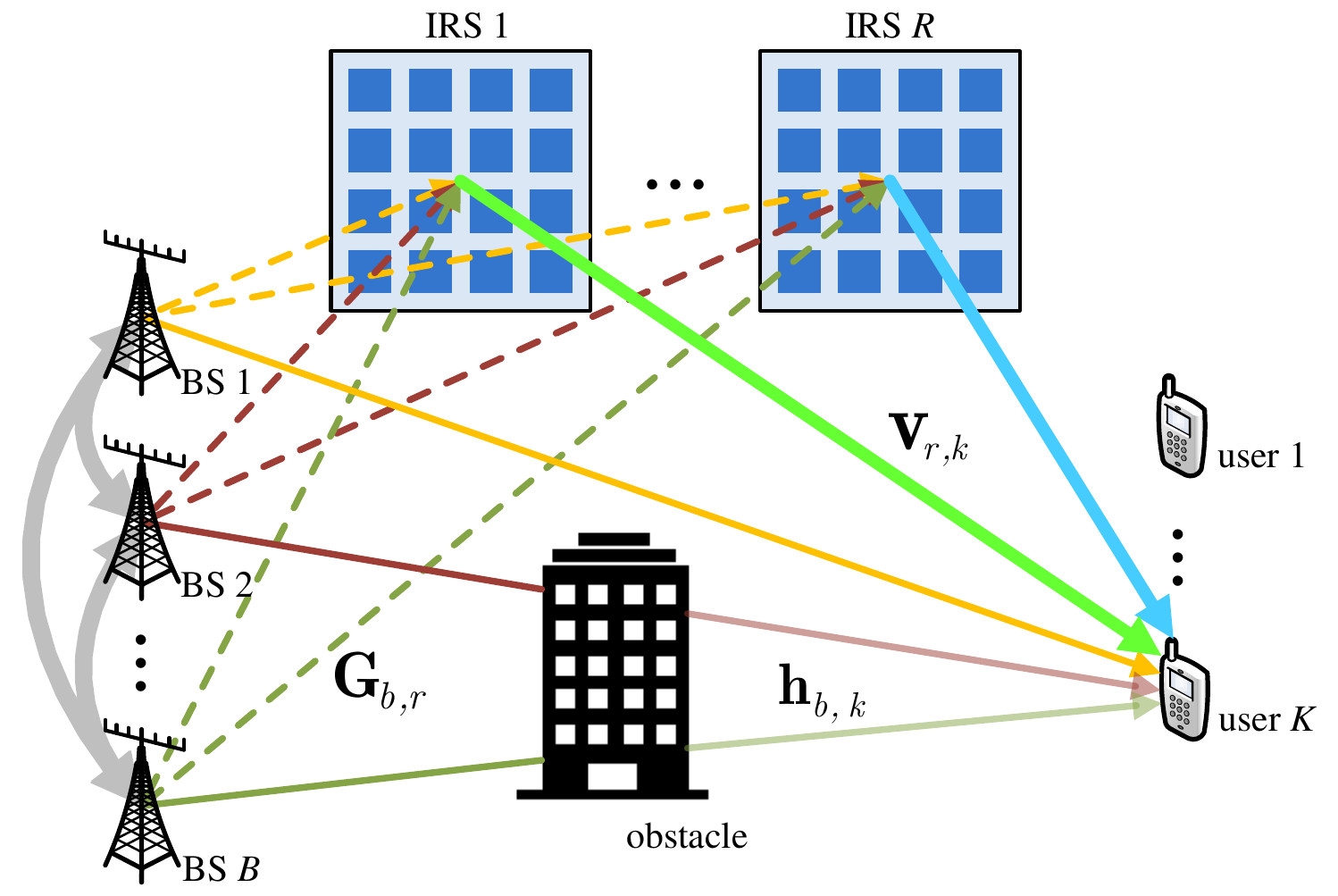}}
    		\caption{ IRS-aided cell-free network.}
    		\label{IRS_aided_systems}
    	\end{center}
    	\vskip -0.2in
    \end{figure}

We consider a downlink RIS-aided cell-free system, as shown in Fig. \ref{IRS_aided_systems}, where a set of BSs $\mathcal{B}=\{1,...,B\}$ and a set of IRSs $\mathcal{R}=\{1,...,R\}$ serve  a set of UEs $\mathcal{K}=\{1,...,K\}$. 
All IRSs are controlled by the BSs by means of wired or wireless control\cite{zhang2020joint}. 
 Let  the number of antennas equipped at  each BS and  UE be  $N_\text{t}$ and $1$, respectively, and the number of reflection elements at each RIS be $N$. 
With the reflection support of IRSs,  the channel between each BS and each UE consists of two parts: the BS-UE link and $R$ BS-RIS-UE links, where each  BS-IRS-UE link is modeled as a concatenation of there components, i.e., the BS-IRS link, IRS phase-shift matrix, and IRS-UE link\cite{guo2020weighted}. Thus, the  equivalent channel between the $b$-th BS and the $k$-th UE is modeled as 
\begin{subequations}
\begin{align}
\widehat{\bf h}_{b,k}^H =&{\bf h}_{b,k}^H+\sum\limits_{r\in \mathcal{R}}  {\bf v}_{r,k}^H  {\bf \Theta}_{r}^H  {\bf G}_{b,r} \\
 =&{\bf h}_{b,k}^H+  {\bm \theta}^H {\bf V}_{k}^H {\bf G}_{b} ,  \label{e_channel} 	
\end{align}
\end{subequations} 
where ${\bf h}_{b,k}  \in \mathbb{C}^{N_t}$, ${\bf v}_{r,k} \in \mathbb{C}^{N}$ and ${\bf G}_{b,r} \in \mathbb{C}^{N \times N_t}$ denote the channel from the  $b$-th BS to the $k$-th UE, from the  $r$-th IRS to the $k$-th UE, and  from the  $b$-th BS to the $r$-th IRS, respectively.  ${\bf \Theta}_{r}= \text{diag}(\theta_{r,1},...,\theta_{r,N}  ) \in \mathbb{C}^{N\times N} $ denotes the 
phase shift   matrix at the $r$-th IRS, where $|\theta_{r,n}|^2=1, \forall r, n$\cite{guo2020weighted}. The equivalent channel can be compactly expressed as \eqref{e_channel} by defining 
 ${\bf V}_{k}=\text{diag}([{\bf v}_{1,k}^T,...,{\bf v}_{R,k}^T])\in \mathbb{C}^{NR \times NR}$, ${\bf G}_b= [{\bf G}_{b,1}^T,...,{\bf G}_{b,R}^T]\in   \mathbb{C}^{NR \times N_t}$, ${\bm \theta}= {\bf \Delta}{\bm 1}_{NR}$ and  ${\bf \Delta}=\text{diag}({\bf \Theta}_1,...,{\bf \Theta}_R)\in   \mathbb{C}^{NR \times NR}$. 
 Then, the IRS constraints can be defined as ${\bm \theta} \in \mathcal{F}$, where $\mathcal{F}$  is the set of $NR$-dimensional vectors of unit-modulus entries. 
Let $s_k \sim \mathcal{CN}(0,1)$ denote the transmitted symbol to  UE $k$. Likewise, let $ {\bf w}_{b}=[{\bf w}_{b,1}^T,..., {\bf w}_{b,K}^T]^T$, where $ {\bf w}_{b,k} \in \mathbb{C}^{N_t}$ is the precoding vector used by BS $b$ for UE $k$. We assume  the  per-BS power constraint  $\sum\nolimits_{k\in \mathcal{K}}  {\left\| {\bf w}_{b,k} \right\|}^2 \le P_b, \forall b$, where $P_b$ denotes the maximum transmit power at  BS $b$. Thus, the received signal at the $k$-th UE is 
\begin{equation}\label{received_signal}
y_k= \sum\limits_{b\in \mathcal{B}} \sum\limits_{j \in \mathcal{K}}  {\widehat {\bf h}}_{b,k}^H {\bf w}_{b,j} s_{j}+z_k,
\end{equation}
where $z_k \sim \mathcal{CN}(0,\delta^2)$ is   average Gaussian noise at UE $k$. Then, the signal-to-interference-plus-noise (SINR) at UE $k$ is 
\begin{equation}\label{SINR}
\varGamma_k =\frac{|\sum\nolimits_{b\in \mathcal{B}} {\widehat {\bf h}_{b,k}^H} {\bf w}_{b,k}  |^2}{ \sum\nolimits_{j\in \mathcal{K}\backslash k}  |\sum\nolimits_{b\in \mathcal{B}} {\widehat  {\bf h}_{b,k}^H} {\bf w}_{b,j}  |^2 +\delta^2}.
\end{equation}

Our objective is to maximize the  WSR of all $K$ UEs by  jointly designing transmitting digital beamformers  and IRS-based analog beamformers, subject to  per-BS transmit power constraints and phase shift constraints. Thus, the centralized WSR maximization
problem is formulated as  
\begin{equation}\label{Problem_main}
\begin{split}
(\text{P}1):~ \mathop {\max }\limits_{{\bm \theta}, {\bf W}}   ~ & R_{s}= \sum\limits_{k\in \mathcal{K}} \omega_k \log(1+\varGamma_k) \\
~~ \text{s.t.}   ~~ &  \sum\limits_{k\in \mathcal{K}}  {\left\| {\bf w}_{b,k} \right\|}^2 \le P_b,~   b\in\mathcal{B};  ~{\bm \theta} \in \mathcal{F}, 
 \end{split}
\end{equation} 
where  ${\bf W}=\{ {\bf w}_{b}| b\in\mathcal{B}\}$.

\section{Decentralized Beamforming Design}

In what follows, we will propose a fully decentralized beamforming scheme to solve problem (P1), where information is exchanged only among neighboring BSs via backhaul signaling and BS-specific beamformers are computed locally by the BSs.

Under decentralized processing,  the IRS-based analog beamformer  computed by each BS should reach consensus. That is, we should guarantee ${\bm \theta}_b={\bm \theta}_{l}$, $b \in \mathcal{B} $ and $l \in \mathcal{B}\backslash b $, where ${\bm \theta}_b$ is the local IRS-based analog beamformer computed at BS $b$. 
On the other hand, the WSR maximization problem (P1) is non-convex w.r.t. ${\bm\theta}_b$ and ${\bf w}_{b}$ due to the coupled variables in the ratio term of WSR in \eqref{Problem_main} and the constant modulus constraints of phase shift vectors.  Therefore, we first transform problem (P1) to a tractable   problem based on  the Lagrangian dual transform and fractional programming theory  \cite{shen2019optimization}.
 By introducing two auxiliary variables ${\bm \gamma} =[\gamma_1,...,\gamma_k]\in \mathbb{R}^K$ and ${\bm \xi}=[\xi_1,...,\xi_K] \in \mathbb{C}^K$,   problem (P1) can be equivalently  rewritten as
\begin{equation}\label{Problem_main2}
\begin{split}
(\text{P} 2):~ \mathop {\min }\limits_{ {\bf \Theta}, {\bf W} }   ~& f({\bf \Theta}, {\bf W}, {\bm \gamma},{\bm \xi})\\   
~~ \text{s.t.}    ~~ &  \sum\limits_{k\in \mathcal{K}}  {\left\| {\bf w}_{b,k} \right\|}^2 \le P_b, ~ b\in\mathcal{B};  \\
 &~{\bm \theta}_b={\bm \theta}_{l}, ~b \in \mathcal{B},~ l \in \mathcal{B}\backslash b; \\
 &~{\bm \theta}_b \in \mathcal{F}, ~b\in\mathcal{B},\\ 
 \end{split}
\end{equation} 
where ${\bf \Theta}=\{{\bm \theta}_b| b\in\mathcal{B}\}$,
 \begin{equation}
 \begin{split}
f({\bf \Theta}, {\bf W}, {\bm \gamma},{\bm \xi}) =&\sum_{k \in \mathcal{K}} \omega_k \Big( \sum_{j\in \mathcal{K}} |\xi_{j}|^2   \bigg| \sum_{b \in \mathcal{B}} {\widehat {\bf h}_{b,j}^H} {\bf w}_{b,k}   \bigg|^2 \\
&- 2\sqrt{1+\gamma_k} \sum_{b \in \mathcal{B}}\text{Re}\big\{ \xi_k^* {\widehat {\bf h}_{b,k}^H} {\bf w}_{b,k} \big\}\\
 &- \log(1+\gamma_k) + \gamma_k + |\xi_k|^2\delta^2 \Big).
 \end{split}
 \end{equation}
 
For the detailed transformation of problem (P2), the reader  is referred to  \cite{shen2019optimization}.  Note that problem (P2) is a  consensus optimization problem w.r.t. ${\bm \theta}_b, \forall b$. Meanwhile,  problem (P2) is a bi-convex optimization problem with fixing ${\bf \Theta}$ and a common practice for solving it is the alternative optimization method. 
To compactly expressed the consensus constraint  in problem (P2), we let $\mathcal{G}= \{\mathcal{B},\mathcal{E} \}$ denote an undirected graph where $\mathcal{B}$ is the BSs and $\mathcal{E}$  includes the connections. Then, the consensus constraint w.r.t. ${\bm \theta}_b, \forall b$, can be reformulated as  ${\bf t}=\sum\nolimits_{b\in \mathcal{B}} {\bf A}_b {\bm \theta}_b=0$, where ${\bf A}_b \in \mathbb{R}^{NR|\mathcal{E}| \times NR}$ is deduced from $\mathcal{G}$\cite{ye2019mobility}. 
  It is still hard to get a global optimal solution of the non-convex problem (P2) under a   consensus constraint.
 To effectively solve  problem (P2), we then utilize the  ADMM method in \cite{ye2019mobility,ye2019decentralized}.
 The augmented Lagrangian for problem (P2) is  
\begin{equation}\label{Lagran}
\begin{split}
 \mathcal{L}({\bf \Theta}, {\bf W}, {\bm \gamma},{\bm \xi},{\bm \lambda}) 
 = & f({\bf \Theta}, {\bf W}, {\bm \gamma},{\bm \xi}) + \sum_{b\in \mathcal{B}}\mu_b \big(\sum_{k\in \mathcal{K}}  {\left\| {\bf w}_{b,k} \right\|}^2- P_b\big)\\
 &+  \sum_{b\in \mathcal{B}} \mathbbm{1}_{\mathcal{F}}({\bm \theta}_b)+\frac{\rho}{2} {\bigg\|  \sum_{b\in \mathcal{B}} {\bf A}_b {\bm \theta}_b+\frac{{\bm \lambda}}{\rho}  \bigg\|}^2,
\end{split}
\end{equation}
where $ {\bm \lambda}$ is a Lagrange multiplier and $\rho>0$, $\mu_b, \forall b$,  is the dual variable introduced for each  per-BS power constraints, $\mathbbm{1}_{\mathcal{F}}( \cdot )$ is   the indicator function of set $\mathcal{F}$ (i.e., $\mathbbm{1}_{\mathcal{F}}({\bm \theta}_b)=0$ if ${\bm \theta}_b\in \mathcal{F}$; otherwise, $\mathbbm{1}_{\mathcal{F}}({\bm \theta}_b)=+\infty$). 

 Since the incremental update method  for decentralized optimization is more communication-efficient than the full CSI exchange method\cite{ye2019decentralized}, we thus utilize   this method  to solve problem (P2). 
Then, variables at BS $b:  = (i_\text{o}+1 \mod B) + 1$ at   the  $i_\text{o}+1$-th iteration can be updated by 
\begin{subequations}\label{Digital}
	\begin{align}	
{\bm \gamma}^{i_\text{o}+1}:=& \arg \mathop{\min}\limits_{{\bm \gamma} }  \mathcal{L}({\bf \Theta}^{i_\text{o}}, {\bf W}^{i_\text{o}}, {\bm \gamma},{\bm \xi}^{i_\text{o}},{\bm \lambda}^{i_\text{o}}),\label{problem_gama}\\
{\bm \xi}^{i_\text{o}+1}:=& \arg \mathop{\min}\limits_{{\bm \xi} } \mathcal{L}({\bf \Theta}^{i_\text{o}}, {\bf W}^{i_\text{o}}, {\bm \gamma}^{i_\text{o}+1},{\bm \xi},{\bm \lambda}^{i_\text{o}}),\label{problem_xi}\\
{\bf w}_{b}^{i_\text{o}+1}:=& \arg \mathop{\min}\limits_{{\bf w}_{b}} \mathcal{L}({\bf \Theta}^{i_\text{o}}, {\bf w}_{b}, \overline{\bf w}_{b}^{i_\text{o}}, {\bm \gamma}^{i_\text{o}+1},{\bm \xi}^{i_\text{o}+1},{\bm \lambda}^{i_\text{o}}),\label{problem_W}\\
	{\bm \theta}_b^{i_\text{o}+1}:=& \arg \mathop{\min}\limits_{{\bm \theta}_b} \mathcal{L}({\bm \theta}_b, \overline{\bm \theta}_b^{i_\text{o}}, {\bf W}^{i_\text{o}+1},   {\bm \gamma}^{i_\text{o}+1},{\bm \xi}^{i_\text{o}+1},{\bm \lambda}^{i_\text{o}}),\label{problem_theta_b}\\
	{\bm \lambda}^{i_\text{o}+1}:=&  {\bm \lambda}^{(i_\text{o} )}+\rho  \sum\limits_{b\in \mathcal{B}} {\bf A}_b {\bm \theta}_b^{i_\text{o}+1}\label{lambda_update2},  
	\end{align}	
\end{subequations} 
where $\overline{\bf w}_{b}={\bf W} \backslash {\bf w}_{b}$ and $\overline{\bm \theta}_b={\bf \Theta } \backslash {\bm \theta}_b$. Note, there are local copies of  ${\bm \gamma}$, ${\bm \xi}$ and ${\bm \lambda}$ in each BS and they are updated locally.

In what follows, we focus on solving   problems  \eqref{problem_gama}-\eqref{problem_theta_b} and the iteration index is dropped to simplify notation.
We first  derive the optimal solutions of problems \eqref{problem_gama}-\eqref{problem_W} in the following proposition. 
\begin{proposition}
The optimal solution ${\bm \gamma}^*$ for problem \eqref{problem_gama} is  
\begin{equation}\label{gamma_update}
\gamma_k^* =\varGamma_k, ~k\in\mathcal{K}.
\end{equation}

The optimal solution ${\bm \xi}^*$ for problem \eqref{problem_xi} is  
\begin{equation}\label{xi_update}
\xi_k^* = \frac{  ( \varphi_{k,k}+  \psi_{k,k} )\sqrt{(1+\gamma_k)\omega_k}  }{ \sum\nolimits_{j\in \mathcal{K} }  |\varphi_{k,j}+  \psi_{k,j}  |^2 +\delta^2},~k\in\mathcal{K},
\end{equation}
where $\psi_{k,j}= \sum\nolimits_{b\in \mathcal{B}} {\bm \theta}_b^H {\bf V}_{k}^H   {\bf G}_{b} {\bf w}_{b,j}$ and  $\varphi_{k,j}= \sum\nolimits_{b\in \mathcal{B}} {{\bf h}_{b,k}^H} {\bf w}_{b,j}$.

Then, the optimal solution ${\bf w}_{b}^*$ for problem \eqref{problem_W} is  
\begin{equation}\label{W_update}
{\bf w}_{b,k}^* = (  {\bf \Phi}_{b} + \mu_b {\bf I}_{N_t}   )^{-1} \big( \sqrt{(1+\gamma_k)\omega_k} \xi_k {\widehat{\bf h}_{b,k}^H}- {\bf \Omega}_{b,k}  \big), \forall k,
\end{equation}
where $ {\bf \Omega}_{b,k} =\sum\nolimits_{j\in \mathcal{K} } |\xi_{j}|^2  {\bf w}_{l,k} {\widehat{\bf h}_{b,j}} ( \varphi_{j,k} +\psi_{j,k} - {\widehat{\bf h}_{b,j}}^H  {\bf w}_{b,k}) $, ${\bf \Phi}_{b} = \sum\nolimits_{j\in \mathcal{K} } |\xi_{j}|^2 {\widehat{\bf h}_{b,j}} {\widehat{\bf h}_{b,j}^H}$, $\mu_b, \forall b$, can be  obtained via bisection methods. 
\end{proposition} 
\begin{IEEEproof}
$\gamma_k^*$ in \eqref{gamma_update}, $\xi_k^*$ in \eqref{xi_update} and ${\bf w}_{b,k}^*$ in \eqref{W_update} can be obtained by respectively solving the  following equations: $ \frac{\partial \mathcal{L}({\bf \Theta}, {\bf W}, {\bm \gamma},{\bm \xi},{\bm \lambda})}{\partial {\gamma}_k}=0$, $ \frac{\partial \mathcal{L}({\bf \Theta}, {\bf W}, {\bm \gamma},{\bm \xi},{\bm \lambda}) }{\partial{\xi}_k}=0$  and $ \frac{\partial \mathcal{L}({\bf \Theta}, {\bf W}, {\bm \gamma},{\bm \xi},{\bm \lambda})}{\partial {\bf w}_{b,k}}=0$. 
\end{IEEEproof}

Then, we will propose an efficient method to solve problem \eqref{problem_theta_b}. To locally optimize ${\bm \theta}_b$ at BS $b$, we first rewrite the augmented Lagrangian in \eqref{Lagran} as 
\begin{equation}\label{Lagran2}
\begin{split}
&\mathcal{L}({\bf \Theta}, {\bf W}, {\bm \gamma},{\bm \xi},{\bm \lambda})\\
 &~~~=  f({\bf \Theta}, {\bf W}, {\bm \gamma},{\bm \xi}) + \sum_{b\in \mathcal{B}} \mathbbm{1}_{\mathcal{F}}({\bm \theta}_b)+\frac{\rho}{2} {\bigg\|   {\bf A}_b {\bm \theta}_b+{\bf t}_b+\frac{{\bm \lambda}}{\rho}  \bigg\|}^2 +C_1,\\
 &~~~={\bm \theta}_b^H {\bf Z} {\bm \theta}_b -2\text{Re}\{ {\bm \theta}_b^H  {\bf q}  \}+ \mathbbm{1}_{\mathcal{F}}({\bm \theta}_b)+C_1+C_2,
\end{split}
\end{equation}
where ${\bf t}_b=\sum_{l\in \mathcal{B}\backslash b} {\bf A}_{l} {\bm \theta}_{l}$, ${\bf Z}=\sum_{k \in \mathcal{K}}\sum_{j \in \mathcal{K}} \omega_k |\xi_{j}|^2 {\bf x}_{j,k}{\bf x}_{j,k}^H+\frac{\rho}{2} {\bf A}_b^H {\bf A}_b$,  ${\bf x}_{j,k}=\text{diag}({\bf v}_{j}^H) {\bf G}_{b}{\bf w}_{b,k}$,
\begin{equation}
\begin{split}
{\bf q}=&  \sum_{k \in \mathcal{K}}\sum_{j \in \mathcal{K}} \omega_k |\xi_{j}|^2 {\bf x}_{j,k} \big( {\widehat{\bf h}_{b,j}}^H  {\bf w}_{b,k}-\varphi_{j,k} -\psi_{j,k}\big) \\
&~+\sum_{k \in \mathcal{K}} \omega_k \sqrt{1+\gamma_k}{\bf x}_{k,k} -\frac{\rho}{2} {\bf A}_b^H\big({\bf t}+\frac{{\bm \lambda}}{\rho}\big),
\end{split}
\end{equation}
and $C_{\{1,2\}}$ are constant terms,  which are not related to ${\bm \theta}_b$. Thus, problem \eqref{problem_theta_b} can be rewritten as 
\begin{equation}\label{Problem_sub2}
\begin{split}
(\text{P} 4):~ \mathop {\min }\limits_{  {\bm \theta}_b} ~    &g_b({\bm \theta}_b) = {\bm \theta}_b^H {\bf Z} {\bm \theta}_b -2\text{Re}\{ {\bm \theta}_b^H  {\bf q}  \}\\ ~~\text{s.t.} ~~ &{\bm \theta}_b \in \mathcal{F}.  
 \end{split}
\end{equation} 

Though the objective function in \eqref{Problem_sub2} is  a simple quadratic function, it is still hard to derive the optimal ${\bm \theta}_b$ with  non-convex unit-modulus constraints. Based on  \cite{wu2017transmit,huang2020learning}, the MM method is an effective way to  solve the non-convex problem (P4). The basic idea is to transform the original problem (P4) into a sequence of majorized subproblems that can be solved with closed-form minimizers. At first, according to lemma 2 in \cite{wu2017transmit},   we can find a valid majorizer of $g_b({\bm \theta}_b)$ at point $ {\bm \theta}_b^{i_\text{i}}\in \mathcal{F}$ given by 
\begin{equation}\label{Majorizer}
g_b({\bm \theta}_b; {\bm \theta}_b^{i_\text{i}})=2\text{Re}\big\{{\bm \theta}_b^H( ({\bf Z}-\zeta{\bf I}){\bm \theta}_b^{i_\text{i}}-{\bf q}) \big\}+C_3,
\end{equation}
where $C_3$ is a constant term, which is not related to ${\bm \theta}_b$, $\zeta$ is the maximum eigenvalue of matrix ${\bf Z}$.
Then, according to the MM method and  utilizing the majorizer in \eqref{Majorizer}, the solution of problem (P4) can be obtained by iteratively solving the following problem 
\begin{equation}\label{Problem_sub3}
\begin{split}
(\text{P} 5):~ \mathop {\min }\limits_{  {\bm \theta}_b} ~ & g_b({\bm \theta}_b; {\bm \theta}_b^{i_\text{i}}), ~~\text{s.t.} ~~ {\bm \theta}_b \in \mathcal{F}.\\ 
 \end{split}
\end{equation} 
The closed-form solution for (P5) is 
\begin{equation}\label{solution_theta}
 {\bm \theta}_b^{i_\text{i}+1}=-\exp \big(j\angle( ({\bf Z}-\zeta{\bf I}){\bm \theta}_b^{i_\text{i}}-{\bf q}) \big).
\end{equation}

The proof of  convergence for the MM  method is similar to \cite{huang2020learning} and omitted here because of the space limitation.

\begin{figure}[t] 
   	\begin{center}
   		\centerline{\includegraphics[width=71mm]{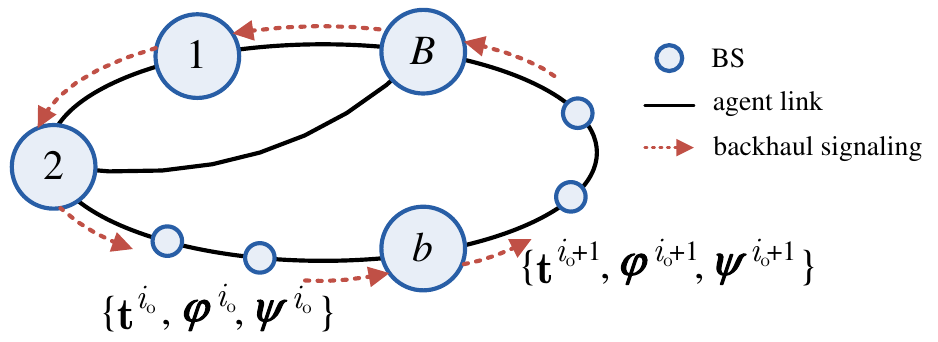}}
   		\caption{Example for the update of Algorithm 1.}
   		\label{Example}
   	\end{center}
   	\vskip -0.2in
   \end{figure}

With  the above analysis, we summarize the proposed fully decentralized beamforming scheme in Algorithm 1.  Fig.  \ref{Example} presents an example of Algorithm 1. Note that, the BSs are activated in a fixed sequencing order and all variables are incrementally updated. 
According to steps 6  and 17 in Algorithm 1, the required backhaul signaling for local variables update at each BS includes one $NR|\mathcal{E}|$-dimensional vector $\bf t$ and two $K\times K$-dimensional matrices, i.e., ${\bm \varphi}$ and ${\bm \psi}$.  The required backhaul signaling is incrementally  updated and transmitted to the next BS after updating ${\bf w}_{b}$  and ${\bm \theta}_b$.  Note that, we do not have to exchange all CSI among BSs in each iteration, and the signaling overhead does not depend on the number of transmit antennas and channels. The total required backhaul signaling of the proposed scheme at each iteration is $B(2K^2+NR|\mathcal{E}|)$ symbols. 
 Thus, the proposed scheme can significantly reduce signaling when compared with full CSI and updated variables exchange among BSs 
 \ \begin{algorithm}[t]\label{ALG2}
   	\caption{Decentralized beamforming  algorithm} 
   	\begin{algorithmic}[1]
   	    \STATE \textbf{Input}: ${\bf h}_{b,k},$, ${\bf V}_{k}$, ${\bf G}_{b}$, ${\bf A}_b$, $\forall b,k$;
         	\STATE \textbf{Output}: ${\bf w}_{b,k}$, $ {\bm \theta}_b$, $\forall b, k$; 
   		\STATE \textbf{Initialize}: ${\bf w}_{b,k}^0 $, ${\bm \theta}_b^0 $, $\forall b, k$,  ${\bm \lambda}^0 $;   
   		\FOR{$i_\text{o}=0,1,...$}
   		\STATE BS $b: =b_{i_\text{o}}= (i_\text{o} \mod B) + 1$ do:
   			 \STATE \textbf{receive} ${\bf t}^{i_\text{o}}$, ${\bm \varphi} ^{i_\text{o}}$ and ${\bm \psi}^{i_\text{o}}$.
   		    \STATE \textbf{update} ${\bf t}_b^{i_\text{o}+1}={\bf t}^{i_\text{o}}-{\bf A}_b {\bm \theta}_b^{i_\text{o}}$;  	
   		 	\STATE \textbf{update} $\overline{ \varphi}_{k,j}^{i_\text{o}+1}=\varphi_{k,j}^{i_\text{o}}-{\bf h}_{b,k}  {\bf w}_{b,j}^{i_\text{o}},  k,j\in\mathcal{K}$; 	
   		 	\STATE \textbf{update} $\overline{ \psi}_{k,j}^{i_\text{o}+1}=\psi_{k,j}^{i_\text{o}}-{\bm \theta}_b^{i_\text{o}H} {\bf V}_{k}^H   {\bf G}_{b} {\bf w}_{b,j}^{i_\text{o}}, k,j\in\mathcal{K}$; 
   		    \STATE \textbf{update} $\gamma_k^{i_\text{o}+1}, \forall k$, using \eqref{gamma_update}; 
   	     	\STATE \textbf{update} $\xi_k^{i_\text{o}+1}, \forall k$, using \eqref{xi_update}; 
   	    	\STATE \textbf{update} ${\bf w}_{b,k}^{i_\text{o}+1}, \forall k$, using \eqref{W_update}; 
   	    	\STATE \textbf{update} ${\bm \theta}_b^{i_\text{o}+1}$ by solving problem \eqref{problem_theta_b}; 
   	    	\STATE \textbf{update} ${\bm \lambda}^{i_\text{o}+1}$ using \eqref{lambda_update2};
   		    \STATE \textbf{update} ${\bf t}^{i_\text{o}+1}  ={\bf t}_b^{i_\text{o}+1}+{\bf A}_b {\bm \theta}_b^{i_\text{o}+1}$;  	    	 	    	
   		 	\STATE \textbf{update} $\varphi_{k,j}^{i_\text{o}+1}=  \overline{\varphi} _{k,j}^{i_\text{o}+1}+{\bf h}_{b,k}{\bf w}_{b,j}^{i_\text{o}+1}, k,j\in\mathcal{K}$; 	
   		 	\STATE \textbf{update} $\psi_{k,j}^{i_\text{o}}=\overline{ \psi}_{k,j}^{i_\text{o}+1}+{\bm \theta}_b^{i_\text{o}+1H} {\bf V}_{k}^H   {\bf G}_{b} {\bf w}_{b,j}^{i_\text{o}+1},k,j\in\mathcal{K}$; 
  			\STATE \textbf{send} ${\bf t}^{i_\text{o}+1}$, ${\bm \varphi} ^{i_\text{o}+1}$ and ${\bm \psi}^{i_\text{o}+1}$ to BS $b_{i_\text{o}+1}$.
  			\STATE \textbf{until} the stopping criterion is met.
  		\ENDFOR 
   	\end{algorithmic} 
   \end{algorithm} 
   
 We then   analyze  the convergence and complexity   of  Algorithm 1 in the following proposition.
\begin{proposition}
 The sequence $({\bf \Theta}^{i_{\text{o}}}, {\bf W}^{i_{\text{o}}}, {\bm \gamma}^{i_{\text{o}}},{\bm \xi}^{i_{\text{o}}},{\bm \lambda}^{i_{\text{o}}})$ generated by Algorithm 1 can converge to a stationary point $({\bf \Theta}^*, {\bf W}^*, {\bm \gamma}^*,{\bm \xi}^*,{\bm \lambda}^*)$ of $\mathcal{L}$, i.e., $0\in \partial \mathcal{L}({\bf \Theta}^*, {\bf W}^*, {\bm \gamma}^*,{\bm \xi}^*,{\bm \lambda}^*)$. When the MM method is used, the main complexity in each iteration of  Algorithm 1 is $\mathcal{O}( KBN_t^3+I_{\text{i}}B(NR)^3) $, where $I_{\text{i}}$ is the number of inner iterations used for MM method. 
\end{proposition} 
\begin{IEEEproof}
According to the general convergence proof for the ADMM method w.r.t. non-convex problems in \cite{wang2019global},  we can find that the objective function of (P2) is continuous and the feasible set $\mathcal{F}$ is bounded, as well as the Lipschitz sub-minimization path  conditions in \cite{wang2019global} are met. Thus,  based on Theorem 2 in  \cite{wang2019global}, we conclude that Algorithm 1 can converge to a stationary point $({\bf \Theta}^*, {\bf W}^*, {\bm \gamma}^*,{\bm \xi}^*,{\bm \lambda}^*)$ of $\mathcal{L}$.  Although the duality gap may  be non-zero, Algorithm 1 still converges and   in general the  dual function at the convergence point is a lower bound of the optimal value of problem (P2). 

In each iteration of  Algorithm 1, the main complexity is the inversion operation  in \eqref{W_update} and finding the maximum eigenvalue of ${\bf Z}$ when using the MM method to solve problem \eqref{problem_theta_b}, which lead to the complexity of $\mathcal{O}(N_t^3)$ and $\mathcal{O}((NR)^3)$, respectively.
\end{IEEEproof}

 The centralized beamforming scheme, where all beamformers are computed by the CPU  of cell-free networks, can be derived following  a similar procedure to \cite{zhang2020joint}.  For more practical implementation  of IRSs, low-resolution discrete phase  shifts should be considered, i.e., $[{\bm \theta}]_i \in \mathcal{F}_2=\{e^{j 2\pi u/2^U} | u=0,...,2^U-1\}$, where the resolution of phase shift is controlled by $U$ bits. Then, according to the nearest point projection in \cite{guo2020weighted},  the solution of (P5) w.r.t. $[{\bm \theta}_b]_i \in \mathcal{F}_2$ can be obtained by solving problem $\angle [{\bm \theta}_b]_i^*=\text{arg} \mathop {\min }\nolimits_{[{\bm \theta}_b]_i \in \mathcal{F}_2} |\angle [{\bm \theta}_b]_i - \angle [{\bm \theta}_b]_i^o|$, where $[{\bm \theta}_b]_i^o$ is the solution with the MM method.

\section{Numerical Results} 
    \begin{figure}[b] 
   	 \vskip 0 in
   	\begin{center}
   		\centerline{\includegraphics[width=70mm]{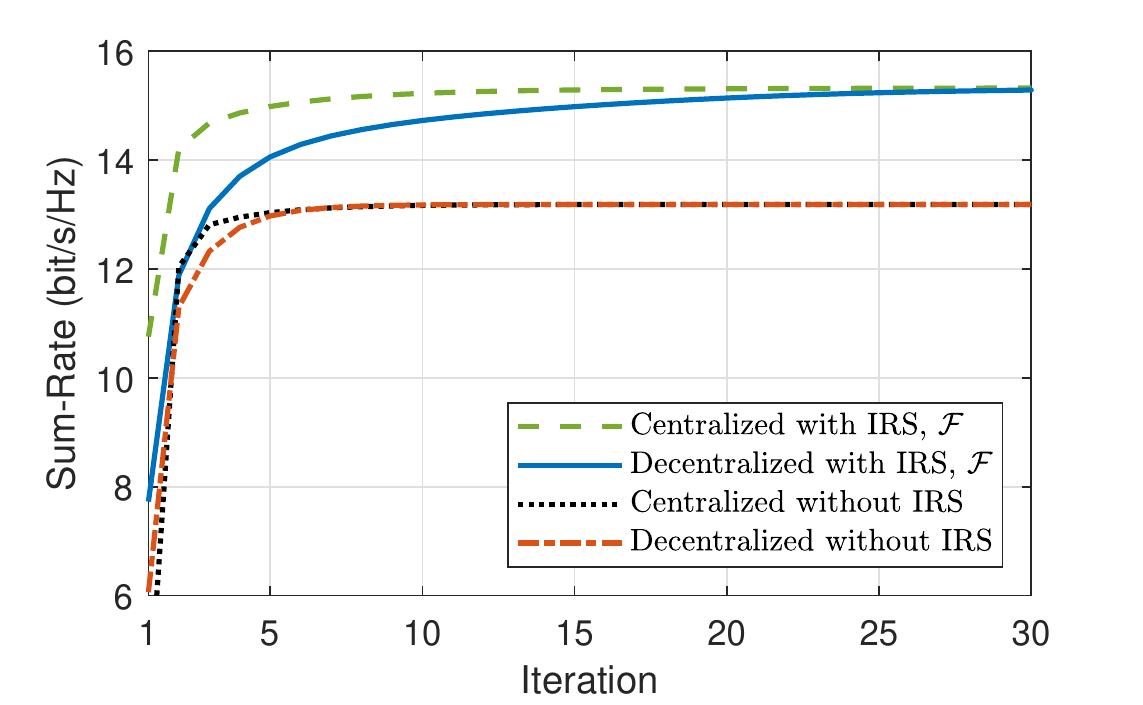}}
   		\caption{Sum-rate vs the number of iterations, with  $N_t=8$, $N=16$, $R=1$.}
   		\label{convergerce}
   	\end{center}
   	\vskip -0.2in
   \end{figure}
     \begin{figure*}[ht] 
        	 \vskip -0.2 in
     	\centering
     	\subfloat[ ]{\hspace*{0 mm}\includegraphics[width =64mm]{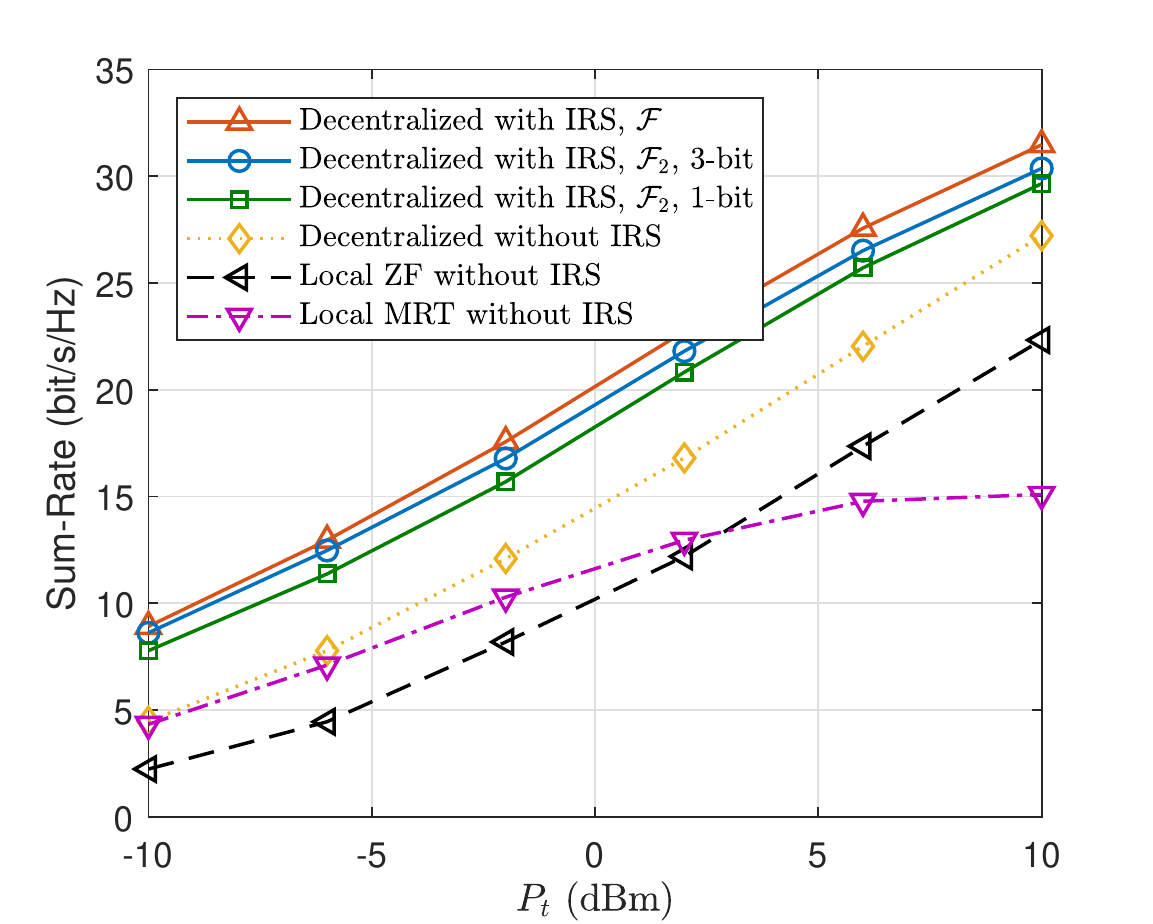}\label{fig_a}}
     	\subfloat[ ]{\hspace*{-3.8mm}\includegraphics[width=64mm]{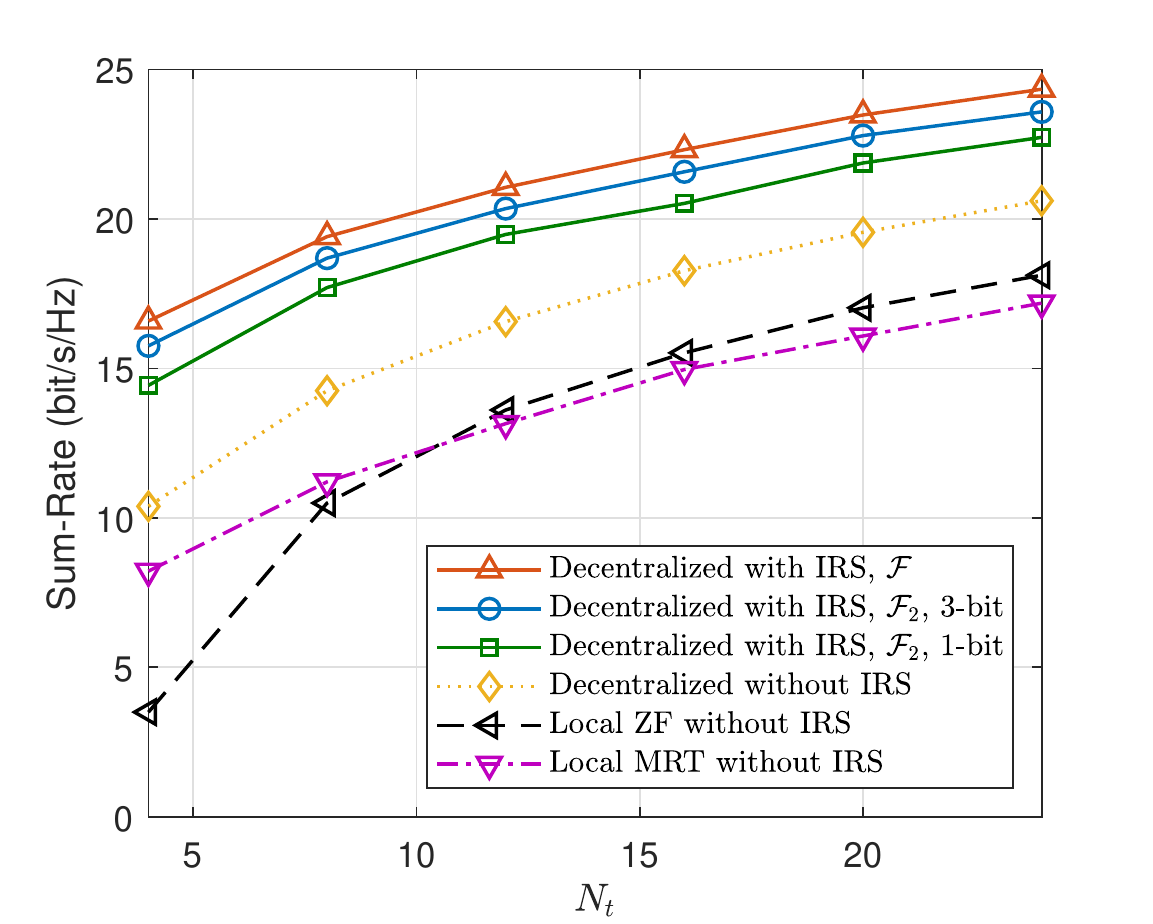}\label{fig_b}}
     	\subfloat[ ]{\hspace*{-3.8mm}\includegraphics[width=64mm]{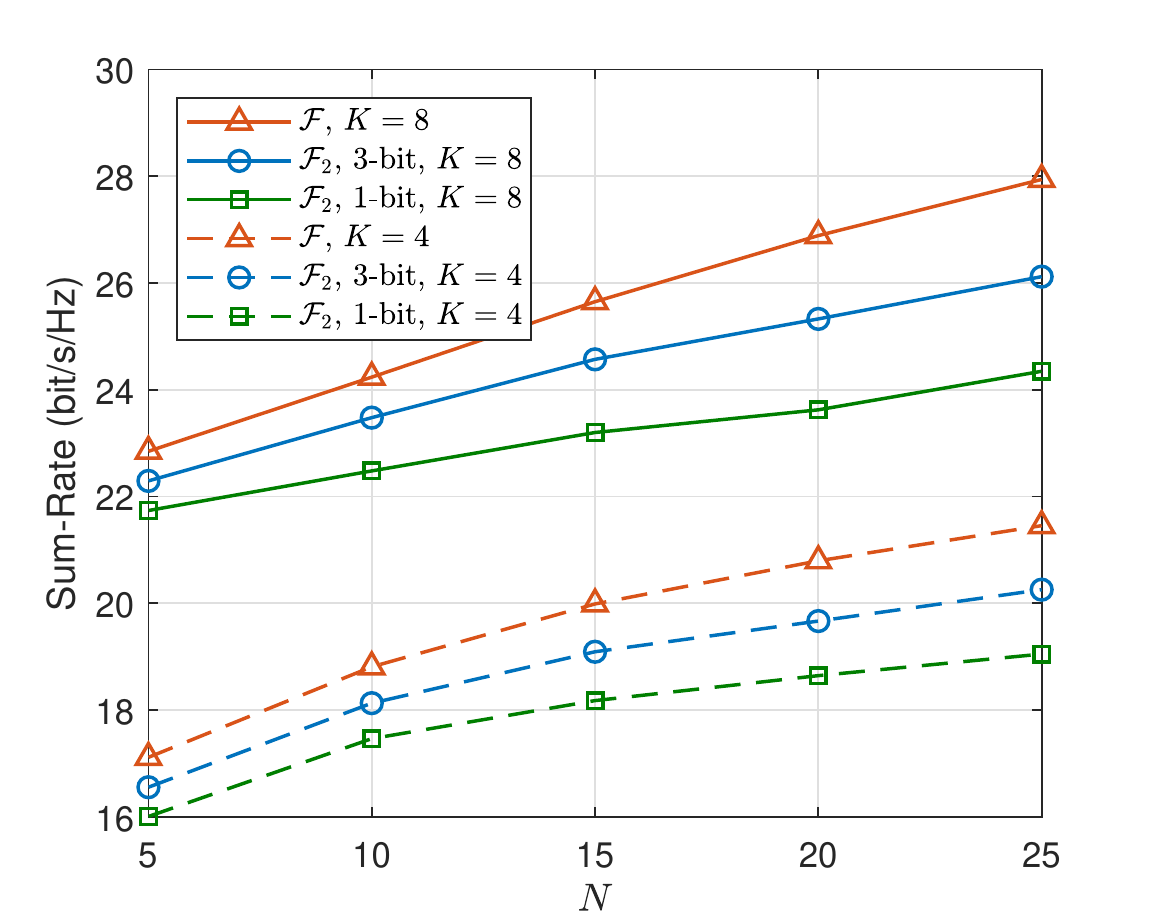}\label{fig_c}}
     	\caption{Sum-rate  vs : (a) $P_t$  with $N_t=8$, $N=16$, $R=3$; (b) $N_t$  with  $N=16$, $R=3$, $P_t=0$; (c) $N$  with $N_t=8$, $R=3$, $P_t=0$.}
     	\label{Fig_all}
     \end{figure*}
In this section, numerical results are provided to evaluate the effectiveness of the proposed algorithm. For large-scale fading, the distance-wavelength-dependent pathloss given as  $PL(d)=\epsilon d^{-\alpha}$, where $d$ is distance, $\epsilon$ is the pathloss at reference distance $1$ m and  $\alpha$ is the pathloss exponent. 
For small-scale fading, we assume that the BS-UE link is non-line-of-sight (NLOS) modeled by Rayleigh fading channels, while the BS-IRS and IRS-UE links are line-of-sight (LOS) modeled by Rician fading channels with Rician factor $0$ dB\cite{guo2020weighted,zhang2020joint}. The bandwidth is  $1$ GHz with central carrier frequency $28$ GHz.  Thus, $\epsilon=-32$ dB and  the pathloss exponents for BS-UE, BS-IRS and IRS-UE links are $3$, $2$, $2$, respectively\cite{rappaport2017overview}. According to \cite{guo2020weighted}, we assume that there is $10$ dB power loss of IRS reflection. The noise power spectral density is  $-174$ dBm/Hz.
 We consider the scenario where  $B=4$ BSs located at $(0,0)$, $(0,2D)$,  $(2D,2D)$ and $(0,2D)$, respectively,    IRSs and UEs are randomly distributed   in a circle centered at $(D,D)$ with radius $0.5D$. Without specific notations, we set $\omega_k=1, \forall k$,  $P_b=P_t=0$ dBm, $\forall b$, $K=4$  and $D=50$ m.

   Fig.~\ref{convergerce} presents the convergence of the proposed algorithm. For the case without IRSs, we see that both decentralized and centralized method can converge very fast (i.e., within $10$ iterations).  For the case with IRSs, since there is a consensus constraint w.r.t. the beamformers of IRSs, it is shown that the convergence rate of the decentralized method is lower than that of centralized methods. For example, the centralized method can converge within $20$ iterations, while the decentralized method can converge within $30$ iterations.  For the cases with and without IRSs, the decentralized method can converge to the same sum-rate as the centralized method. The result verifies the effectiveness of the proposed    method.
    
    Fig.~\ref{Fig_all}  shows the achievable sum-rate of various beamforming methods versus transmit power $P_t$, the number of transmit antennas $N_t$, the number of UEs $K$ and the number of reflection elements $N$. 
    Fig.~\ref{fig_a} shows that the sum-rate of proposed decentralized beamforming methods increases with $P_t$, and outperforms  that of local ZF and MRT methods.  The system with the aid of IRSs can achieve a higher sum-rate than that without IRSs. Moreover, the low-resolution phase shifts  suffer  acceptable performance loss. For instance, the system with      ``3-bits'' phase shifts suffers the sum-rate loss of  $4 \% $   of that with continuous phase shifts.     
    From Fig.~\ref{fig_b}, we can see that the sum-rate of all beamforming methods increases with  $N_t$. The sum-rate of proposed decentralized beamforming methods still outperforms  that of local ZF and MRT methods. Fig.~\ref{fig_c} shows the sum-rate of proposed decentralized beamforming  methods  versus $K$ and $N$.  We see that the sum-rate increases with $K$ and $N$. However, as $N$ and $K$ increases, the sum-rate gap between low-resolution phase shifts and continuous phase shifts increases. This reveals that  the  efficient quantization level of phase shifts are related to $N$ and $K$.   
  \section{Conclusions}
   
   We have developed   a decentralized design framework for cooperative beamforming  in IRS-aided cell-free networks. Based on incremental ADMM methods, a fully decentralized beamforming scheme has been proposed to   locally update  beamformers, in which both transmitting digital beamformers and IRS-based analog beamformers are jointly optimized. The convergence of the proposed method   has been proven and the main complexity has been analyzed.  Results show that the  proposed method for the cases with and without IRSs can achieve better performance than existing decentralized methods (i.e., local MRT and ZF methods). Moreover,  it has been shown that  IRS-aided cell-free networks outperform conventional cell-free networks, and that the system sum-rate increases with the number of transmit antennas and the size of IRSs.    
   Finally, we have seen that, to achieve acceptable performance loss,  the quantization level of phase shifts is related to  the size of IRSs and  the number of transmit antennas. 
   
\bibliography{ref}
\bibliographystyle{IEEEtran}

\end{document}